\documentclass[preprint]{aastex}

\slugcomment{Not to appear in Nonlearned J., 45.}

\shorttitle{Precession in the inner jet of 3C\,345}
\shortauthors{A. Caproni et al.}

\begin{document}


\slugcomment{Accepted by The Astrophysical Journal}
\shorttitle{Precession in the inner jet of 3C\,345}
\shortauthors{A. Caproni et al.}

\title{Precession in the inner jet of 3C\,345}


\author{A. Caproni and Z. Abraham}
\affil{Instituto de Astronomia, Geof\'\i sica e Ci\^encias Atmosf\'ericas, Universidade de S\~ao Paulo,
              R. do Mat\~ao 1226, Cidade Universit\'aria, CEP 05508-900, S\~ao Paulo, SP, Brazil}
\email{acaproni@astro.iag.usp.br}







\begin{abstract}
   VLBI observations have shown that the parsec-jet of 3C 345 is formed by several components, ejected 
   from the core with superluminal velocities and travelling along bent trajectories on the plane of 
   the sky. We interpret the differences in velocity and position angle among the different features 
   at formation time as the result of parsec-scale precession of the relativistic jet and calculate 
   the aperture angle of the precession cone, the angle between the cone axis and the line of sight 
   and the Lorentz factor associated with the jet bulk motion. We assumed a precession period of 10.1 yr, 
   which is one of the {\it B}-band light curve long-term periods reported in the literature. We propose 
   that boosting of the underlying jet emission, which is time-dependent due to precession, is responsible 
   for this long-term optical variability. Jet precession with periods of several years can be produced 
   in super-massive black hole binary systems, when the secondary black hole is in an orbit non-coplanar 
   with the primary accretion disk, inducing torques in the inner parts of the disk. Assuming that 
   this mechanism is responsible for the jet precession in 3C\,345, we estimate upper and lower limits 
   for the masses of the two black holes, as well as their mean separation. We found a correlation 
   between the formation of jet components and the occurrence of strong optical flares, as well as a 
   very strong anti-correlation between the intensity of these flares and the time required for the 
   components to reach the maximum flux density at radio frequencies.
\end{abstract}




\section{Introduction}

   The quasar 3C\,345 (z=0.5928; \citealt{mar96}), also known as 1641+399 or 4C\,39.48, 
   was optically identified by \citet{goki65}, and it is now known to be the nucleus 
   of an elliptical E3 galaxy \citep{kir99}. Strong variability (typical of an OVV 
   object) is found in its optical light curve, and periodicities of about 5 and 11 yr have been 
   reported \citep{web88,kidg89,zha98,zha00}, although it has also been suggested that 
   this variability may originate from non-linear and non-stationary stochastic processes \citep{vio91}.

   The radio flux density has been monitored at several frequencies \citep{wal91,all96,ter98}. 
   The continuum spectrum is flat up to 10 GHz and gets steeper towards higher frequencies, with 
   spectral index ranging from -0.9 (between 10$^{11.9}$ and 10$^{13.2}$ Hz) to -1.4 (between 
   10$^{13.2}$ and 10$^{15.4}$ Hz) \citep{bre86}. Outbursts have been detected from centimeter to infrared 
   wavelengths \citep{bre86,ste96}, besides time variable linear polarization at radio and optical 
   frequencies \citep{most81,bre86,bro94}. In the optical regime, the degree of polarization 
   decreases monotonically toward shorter wavelengths \citep{sit85,smi86,mea88,deDie94} and 
   in the X-ray range, the source is weak and possibly variable \citep{halp82,maki89,wowi90}.

   VLA observations showed a faint halo around a bright core and an extended kiloparsec-jet 
   \citep{kol89}. At parsec-scales, 3C\,345 exhibits a stationary core \citep{bar86,tan90} 
   and a jet with superluminal components, which travel apparently along curved paths with 
   variable velocities \citep{unw83,bir86,zen95,ros00}. These characteristics had been 
   interpreted as a result of helical motion of the components along the jet, either in 
   a pure phenomenological model \citep{qia96} or as a consequence of HD or MHD instabilities 
   (e.g., \citealt{koch85,came86,hard87,qia91,qia92,cakr92,ste95,hard00}). 

   An alternative approach is to assume that the helicoidal appearance of the jet is a consequence 
   of precession. In this case, and considering that the bulk velocity is very large, 
   each plasma element in the jet would move in an almost straight trajectory, defined by the 
   jet direction when this element was ejected. From the observational point of view, superluminal 
   components can be interpreted as these plasma elements, even if they are the results of shock 
   waves propagating along the jet; their velocity should not be very different from the bulk 
   velocity (e.g., \citealt{mage85}) and, at least close to their formation epoch, 
   they should reflect the precessing jet direction. Curved trajectories, as found by 
   \citet{zen95} and \citet{loba96} in 3C\,345, can be due to several reasons: 
   jet instabilities (e.g., \citealt{hard00}), influence of the external environment 
   (one of the mechanisms proposed by \citealt{wal87} to explain the bending of the jet in 3C\,120), 
   or even the superposition with other components formed at different epochs and moving with 
   different superluminal velocities. In any case, the parameters of the precession model can 
   be completely determined by the superluminal velocities and position angles of the jet 
   components close to their formation epoch. 

   The parsec-jet precession scenario has already been applied to 3C\,279, 3C\,273 and OJ\,287 
   \citep{abca98,abro99,abra00}, where the velocity variations among components were attributed 
   to differences in the angle between their trajectories and the line of sight. Besides, it was 
   found that the periodic outbursts seen in the optical light curves of 3C\,279 and OJ\,287 could 
   be due to changes in the boosting parameter \citep{abca98,abra00}. Although this model explained 
   well the behavior of these objects, the precession parameters were obtained from a limited set 
   of data and should probably be revised to include new and more precise data \citep{hom01,weh01}.
   We will show in this paper that the variations in the ejection angles and velocities of the 
   jet components of 3C\,345 can also be interpreted as the result of parsec-jet precession. 
   Furthermore, we show that the presence of long-term optical variability can be attributed to 
   variable boosting of the non-thermal radiation from the underlying jet. 

   In Sect.2, we describe the precession model used in this work. In Sect.3, we present the precession 
   model parameters for 3C\,345, as well as the influence of precession in the absolute core position 
   shifts due to opacity effects. In Sect.4, we discuss, in the framework of the precession model, other 
   observed characteristics, such as the relation between optical flares and the origin of superluminal 
   components, {\it B}-band long-term periodic variability, polarimetric observations and dispersion in the 
   apparent angular size of the components. Furthermore, 
   orbital parameters of a possible super-massive black hole binary system in the nucleus of 3C\,345 are 
   estimated based on the precession model, with additional constrains from optical continuum and spectroscopic 
   data. Finally, conclusions are presented in Sect.5. 

\section{Precession model}

   Let us consider a relativistic jet with bulk velocity $\beta$, precessing with a constant angular 
   velocity $\omega$ and period $P$ around an axis, forming a conical surface with aperture angle $\Omega$. 
   The cone axis forms an angle $\phi_{\mathrm{0}}$ with the line of sight and presents a projected angle 
   $\eta_{\mathrm{0}}$ on the plane of the sky. The instantaneous position of the jet is represented by 
   the angles $\phi$ and $\eta$. Their dependence upon the time $t^\prime$, measured in the comoving 
   frame, can be expressed as \citep{abro99,abra00}:

   \begin{eqnarray}
      \eta(t^\prime) = \arctan(\frac{y}{x}) 
   \end{eqnarray}
   \begin{eqnarray}
      \phi(t^\prime) = \arcsin(\sqrt{x^2+y^2}) 
   \end{eqnarray}
   \\with 

   \begin{eqnarray}
      x = (\cos\Omega\sin\phi_{\mathrm{0}}+\sin\Omega\cos\phi_{\mathrm{0}}\sin\omega t^\prime)\cos\eta_{\mathrm{0}}-\sin\Omega\cos\omega t^\prime\sin\eta_{\mathrm{0}}
   \end{eqnarray}
   \begin{eqnarray}
      y = (\cos\Omega\sin\phi_{\mathrm{0}}+\sin\Omega\cos\phi_{\mathrm{0}}\sin\omega t^\prime)\sin\eta_{\mathrm{0}}+\sin\Omega\cos\omega t^\prime\cos\eta_{\mathrm{0}}
   \end{eqnarray}

   The apparent velocity $\beta_{\mathrm{obs}}$ (in units of light speed $c$) is related to the proper motion 
   of the jet components $\mu$ through:

   \begin{eqnarray}
      \beta_{\mathrm{obs}} = \frac{D_{\mathrm{L}}}{(1+z)}\frac{\mu}{c} 
   \end{eqnarray}
   \\where $z$ is the redshift and $D_{\mathrm{L}}$ is the luminosity distance, defined as \citep{car92}:

   \begin{eqnarray}
      D_{\mathrm{L}} = \frac{c(1+z)}{H_{0}}E(\Omega_{\mathrm{M}},\Omega_{\mathrm{\Lambda}},z)
   \end{eqnarray}
   \\where $H_{0}$ is the Hubble constant and $E$ is a function that depends on $z$ and the 
   dimensionless density parameters $\Omega_{\mathrm{M}}$ and $\Omega_{\mathrm{\Lambda}}$. 
   Throughout the paper, it will be adopted $H_{0}=71$ km s$^{-1}$ Mpc$^{-1}$, $\Omega_{\mathrm{M}}=0.27$ 
   and $\Omega_{\mathrm{\Lambda}}=0.73$, as derived from the recent {\it WMAP} results \citep{ben03}. 
   In this case, since $\Omega_{\mathrm{M}}+\Omega_{\mathrm{\Lambda}}=1$  
   the function $E$ is given by \citep{car92}:

   \begin{eqnarray}
      E(\Omega_{\mathrm{M}},\Omega_{\mathrm{\Lambda}},z) = \int^{z}_{0}\frac{dz^\prime}{\sqrt{(1+z^\prime)^2(1+\Omega_{\mathrm{M}}z^\prime)-z^\prime(2+z^\prime)\Omega_{\mathrm{\Lambda}}}} 
   \end{eqnarray}

   For the chosen cosmology, $D_{\mathrm{L}}=2.47h^{-1}$ Gpc, with $h=H_{0}/100$, from which 1 mas yr$^{-1}$ corresponds to 
   an apparent velocity of 24.5$h^{-1}c$. 

   The viewing angle $\phi$ is not measured directly, but it is related to $\beta_{\mathrm{obs}}$ by:

    \begin{eqnarray}
      \beta_{\mathrm{obs}} = \frac{\beta\sin\lbrack\phi(t^\prime)\rbrack}{1-\beta\cos\lbrack\phi(t^\prime)\rbrack} 
   \end{eqnarray}
 
   The elapsed time between two events in the observer and comoving frameworks ($\Delta t$ and $\Delta t^\prime$ 
   respectively) are related by the Doppler factor $\delta$ as: 

   \begin{eqnarray}
      \Delta t^\prime = \frac{\delta(\phi,\gamma)}{(1+z)}\Delta t 
   \end{eqnarray}
   \\with

   \begin{eqnarray}
      \delta(\phi,\gamma) = \gamma^{-1}\lbrace1-\beta\cos\lbrack\phi(t^\prime)\rbrack\rbrace^{-1} 
   \end{eqnarray}
   \\and 

   \begin{eqnarray}
      \gamma = (1-\beta^2)^{-1/2} 
   \end{eqnarray}

   Thus, given $\Omega$, $\phi_{\mathrm{0}}$, $\eta_{\mathrm{0}}$, $\beta$ and $P$, we could 
   predict at any time the observed properties of a precessing jet.
   
   The flux density in the optically thin regime will be boosted according to the expression: 
   
   \begin{eqnarray}
      S_{\mathrm{j}}(\nu) = S_{\mathrm{j}}^\prime(\nu)\delta(\phi,\gamma)^{p+\alpha} 
   \end{eqnarray}
   \\where $\alpha$ is the spectral index ($S_{\nu}\propto \nu^{-\alpha}$), $p=2$ for a continuous jet and 
   $p=3$ for discrete features \citep{blko79,libl85}.

\section{Determination of the precession parameters}

   In this work we used the VLBI data of 3C\,345  (core-component distance $r$, position angle 
   on the plane of the sky $\eta$ and flux density $S_{\nu}$) found in the literature, between 5 and 
   22 GHz, covering almost 20 yr of monitoring \citep{bir86,unwe92,unw94,bro94,zen95,lep95,loba96,ros00}.
   We adopted the emergence epochs for the identified components given in the references, 
   except for C7, for which we ruled out those observations with not clear identification (e.g., C76 in 
   \citealt{unwe92} and \citealt{unw94}). As there are no observations of the older components C2 and C3 close 
   to their ejection epoch 
   and C9 has only one reported observation \citep{ros00}, we did not use them in our fitting.
   In order to estimate the apparent proper motion $\mu$ and position 
   angles of the components, it was assumed quasi-ballistic motions in the inner region of the quasar 
   ($r<1.0$ mas), since the data show that the trajectories are bent beyond this distance. It is important 
   to emphasize that some components, such as C4 and C5, present variable proper motions and/or position 
   angles even for $r<1.0$ mas (e.g., \citealt{loba96}); in those cases, we have considered only observations 
   taken before the occurrence of significant changes in those quantities. We should note that all the 
   superluminal features became optically thin at distances smaller than 1 mas, since this fact is important 
   for the discussion in Sect. 4.1.
 
   As we used data obtained at different frequencies, it was necessary to take into account 
   opacity effects on the determination of the absolute core position, which introduce frequency-dependent 
   shifts in the core-component distances and proper motions. In the case of a precessing jet, 
   these corrections are time dependent and its magnitude calculated in Appendix A. The parameters 
   involved in this calculation are the integrated synchrotron luminosity $L_{\mathrm{syn}}$, 
   the ratio between upper and lower limits in the relativistic particle energy distribution 
   $\gamma_{\mathrm{max}}/\gamma_{\mathrm{min}}$, the intrinsic jet aperture angle $\psi^\prime$, and 
   a constant parameter $k_{\mathrm{e}}$. Following \citet{loba98}, we assumed 
   $\gamma_{\mathrm{max}}/\gamma_{\mathrm{min}}=100$,  $k_{\mathrm{e}}=1$ and a constant value 
   of $1.58\times10^{45}$ erg s$^{-1}$ for $L_{\mathrm{syn}}$, even though this quantity could 
   be time variable. Finally, we chose $\psi^\prime=1\degr$ based on the observed angular size 
   of the jet components as a function of their distances from the core.

   Among the parameters discussed in Sect.2, we  constrained the precession period $P$ 
   using the {\it B}-band photometric data. In fact, \citet{zha98} reported a period of $10.1~\pm~0.8$ yr 
   in the light curve of 3C\,345 after application of Jurkevich $V_{m}^2$ test \citep{jurk71}. We assumed 
   that this periodicity is due to variable boosting of the  jet emission as the Doppler factor varies with 
   precession (equation [12]). 
   
   We adopted an unique  Lorentz factor  for the bulk motion of all components, compatible with the velocity 
   of the fastest component ($\approx8.7 h^{-1}$ c) given by $\gamma_{min}=(1+\beta_{\mathrm{obs}}^2)^{1/2}$ 
   ($\gamma \geq 12.3$ for $h=0.71$). After fixing $\gamma$ close to its lower limit, we selected the parameters 
   $\Omega$, $\phi_{\mathrm{0}}$ and $\eta_{\mathrm{0}}$ that fitted the apparent velocities and position angles 
   of the jet components using equations (1)-(4), with $t^\prime$ as implicit variable. Then, we checked the 
   behaviour of $h\beta_{\mathrm{obs}}$ and $\eta$ as functions of time through equations (1)-(5), assuming that 
   the position angle of the jet component represents the jet position at the epoch when the component was formed. 
   This procedure was repeated until a good fitting was obtained. For each set of precession parameters, we applied 
   the core opacity shifts to the observational data, which introduced refinements in the kinematic and model
   parameters. The reduced chi-square value for both $\beta(t)$ and $\eta(t)$, for six superluminal components and
   three parameters is about 3, most of it due to uncertainties in the position angle $\eta$.

   In Fig. 1, we present the difference in core position $\Delta r_{\mathrm{core}}$ between 5 and 22 GHz (the 
   frequency range used in our work) as a function of time, calculated with the precession parameters. The mean 
   value obtained for $\Delta r_{\mathrm{core}}$ is 0.26 mas, while its lower and upper limits are respectively 
   0.21 and 0.31 mas. \citet{loba96} found an average offset of $0.328~\pm~0.020$ mas between 1992.45 and 1993.72, 
   in good agreement with our estimate of 0.30 mas for the same interval.

\begin{figure}
      {\plotone{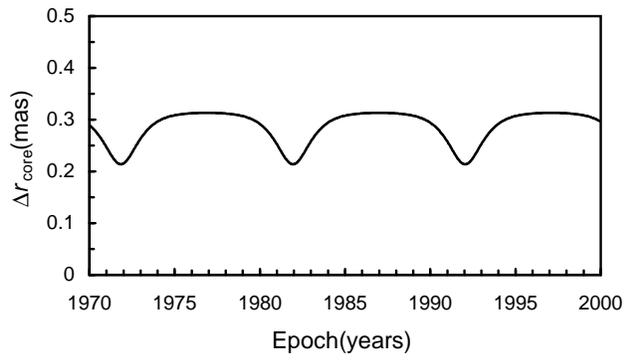}}
      \caption{Time behaviour due to jet precession of the difference between core position obtained at 
               5 and 22 GHz. 
             }
         \label{core shift opacity}
   \end{figure}

   \placefigure{core shift opacity}

   The kinematic parameters of the jet components corrected by opacity effects are presented in Table 1. 
   The quoted errors correspond to the range for which a reasonable fitting for the data could still be 
   found. 
   \placetable{tbl-1}
    
   The precession parameters that best fitted the data are given in Table 2, while Fig. 2 presents the model 
   fitting in the ($\eta$, $h\beta_{\mathrm{obs}}$), ($t$, $h\beta_{\mathrm{obs}}$), ($t$, $\eta$), 
   ($t$, $\delta$) and ($t$, $\psi$) planes. Lower limits for $\delta$ (full triangles) calculated from X-ray 
   observations \citep{unw83,unw97} are presented in the same figure. The Doppler factors predicted by our 
   model are always above the lower limits imposed by the X-ray observations. In fact, although other 
   combinations of the precession parameters $\phi_{\mathrm{0}}$, $\eta_0$, $\Omega$ and $\gamma$ also 
   fitted the kinematic data, they resulted in Doppler factors incompatible with the lower limits imposed 
   by the observed X-ray fluxes. 
   
   Another consequence of the precession model is a possible dispersion in the component angular sizes.
   As mentioned before, the observed  sizes $\psi$ depend on the angle between the jet and the line of sight 
   (equation [A2]). Observations show a dispersion $11\degr \leq\psi\leq 45\degr $ with a mean value of $<\psi>=28\degr$, 
   at least for distances $r<1$ mas, where we assumed ballistic motion. These observations, together with 
   the modeled time behaviour of the apparent jet aperture angle (assuming  an intrinsic aperture $\psi^\prime=1\degr$) 
   are presented in Fig. 2e. The precession model predicts $14\degr \leq\psi\leq 42\degr$ and $<\psi>=28\degr$ 
   and, except for C3a, there is good agreement between the individual observations and the model, which seems 
   to indicate that the jet components have similar intrinsic sizes. If we assume that the time behaviour of 
   $\psi$ is fully described by precession effects, there are several alternatives to explain the discrepancy 
   between the calculated and predicted values for C3a: its intrinsic size is larger than $1\degr$, the 
   estimated value of its apparent size is uncertain due to its low flux density or to its possible relation 
   with C4 \citep{loba96}, or finally because few observations are available for $r<1$ mas.
    
\begin{figure}
      {\plotone{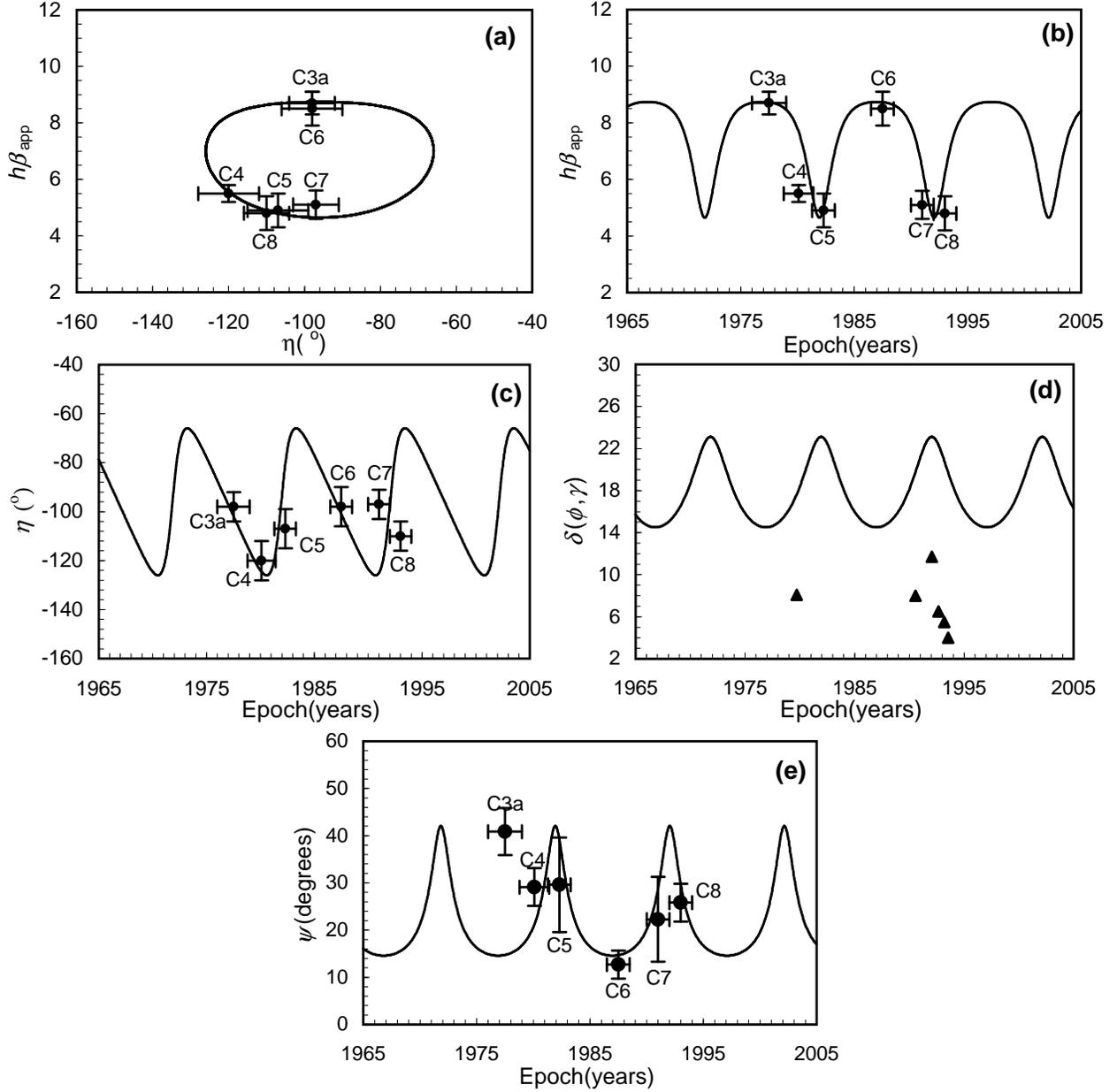}}
      \caption{Precession model applied to the parsec-scale jet of 3C\,345 with model parameters listed in Table 2. 
                 {\bf (a) - (e):} Continuous lines are model predictions on the planes 
				 ($\eta$, $h\beta_{\mathrm{obs}}$), 
                 ($t$, $h\beta_{\mathrm{obs}}$), ($t$, $\eta$), ($t$, $\delta$) and ($t$, $\psi$). Full circles 
				 and triangles represent, respectively, 
				 observations and lower 
                 limits for the Doppler boosting factors, calculated from X-ray observations \citep{unw83,unw97}.
              }
         \label{precession model}
   \end{figure}

   \placetable{tbl-2}
   \placefigure{precession model}

\section{Discussion}

   General results obtained from previous sections are discussed below.


\subsection{The optical light curve}

   In Fig. 3, we show the {\it B}-band light curve of 3C\,345 at the observer's reference frame
   \citep{zha98}. We also show, as a continuous line, the boosted emission of the underlying jet, 
   calculated from equation (12) with $ \alpha_{\mathrm{opt}}=1.66$ \citep{hag96}, $p=2$, the Doppler 
   factor obtained from the precession model and the flux density in the comoving reference frame 
   $S_{\mathrm{j}}^\prime=7$ nJy, which gives the right values for the observed flux density at 
   epochs 1971-1972 and 1990-1991. The inferred value for the underlying jet flux density is an 
   upper limit, since the total luminosity in the optical band is probably due to the superposition 
   of other processes, such as emission from the accretion disk and from the superluminal components.    
   The existence of a weak underlying jet was already postulated in 3C\,279 and OJ\,287 \citep{abca98,abra00}.

\begin{figure}
      {\plotone{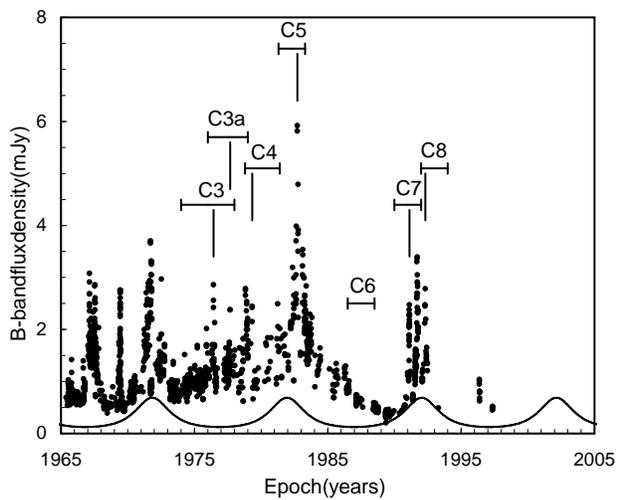}}
      \caption{{\it B}-band light curve of 3C\,345 (full circles). Solid line represents the expected contribution of the 
               underlying jet. The uncertainty in the emergence epoch of each jet component is represented by the 
			   horizontal lines under the component label
               (see Table 1), while the vertical lines mark the flares that could be associated with  these 
               components.}
         \label{B-band light curve}
   \end{figure}

   \placefigure{B-band light curve}

   From the light curve, and whenever observations were available, we were able to associate the emergence 
   of jet components to the occurrence of flares in the optical band\footnote{We have considered as optical 
   flares short-time variations (between 5 and 90 days) that introduce changes in the flux density higher 
   than 0.5 mJy.}, even though for the older jet components uncertainties in $t_{\mathrm{0}}$ lead to a 
   non-unique association. The emergence time of the superluminal features and their associated flares
   are explicitly marked in Fig. 3. There we can also see that only flares stronger than a certain limit 
   (about 1.5 mJy) produced detectable superluminal components. 
   
   It is not clear whether these optical flares are related to short-lived stages of shock evolution 
   (e.g. \citealt{mage85}), or a consequence of some instability produced in the accretion disk \citep{rom00}. 
   An unexpected relation between the flare flux density at optical wavelengths and the elapsed time between
   the epochs of superluminal component ejection and occurrence of maximum flux density at radio frequencies 
   gave us some insight on the origin of these flares. Considering only the interval for which the movement 
   of the jet components is quasi-ballistic, and defining $\Delta T_{\mathrm{max}}(\nu)$ as the the elapsed 
   time between the epoch when the components reach their maximum intensity $t_{\mathrm{max}}(\nu)$ at the 
   observed radio frequency $\nu$ and their formation epoch ($t_{\mathrm{0}}$), we obtain:

   \begin{eqnarray}
      \Delta T_{\mathrm{max}}(\nu) = t_{\mathrm{max}}(\nu) - t_{\mathrm{0}} 
   \end{eqnarray}

   The values of $\Delta T_{\mathrm{max}}$ for C3, C3a, C4, C5 and C7, obtained from the flux densities 
   given in \citet{loba96}, are presented in Table 3. We ignored C2 because the maximum occurred before 
   the first observation, and C8 and C9 because few observations are available. We calculated $\Delta T_{\mathrm{max}}$ 
   at 22 GHz for all components except for C3 and C3a, for which we used data at 10.7 GHz, because
   they did not have good time coverage at 22 GHz. The error introduced by the use of
   a different frequency did not change appreciably our results.

   \placetable{tbl-3}

   In Fig. 4a, we show the {\it B}-band flux densities $S_{\mathrm{B}}$ of the flares associated with the jet components 
   as a function of $\Delta T_{\mathrm{max}}(\nu)$. This plot shows a clear and unexpected anti-correlation between these 
   quantities ($R=-0.97$, where $R$ is the correlation coefficient obtained from a power-law fitting). This kind of 
   correlation was not seen when the maximum radio flux density was considered.

\begin{figure}
      {\plotone{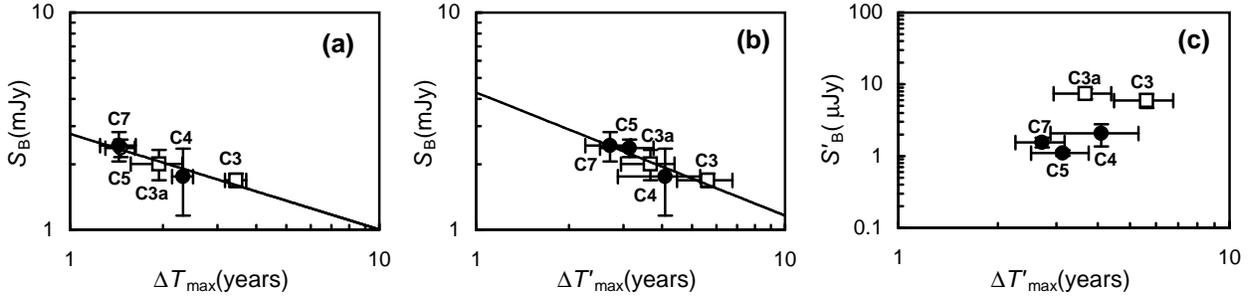}}
      \caption{Comparison between radio and optical properties of the superluminal components and flares. Filled symbols 
               represent jet components for which 22 GHz radio observations were used, 
			   while open symbols correspond to 10.7 GHz observations. {\bf (a):} {\it B}-band 
               flux density of the flares associated with the superluminal features as a 
               function of the elapsed time between
               the epochs of  component ejection and occurrence of maximum flux density at radio frequencies. 
			   {\bf (b):} The same as (a), but as a function of the elapsed time in the comoving
			   jet reference frame. {\bf (c):} The same as (b), but considering boosting of the optical flux
			   density.
               The solid line in each plot corresponds to the minimum least squares fitting using power laws. 
              }
         \label{correlations between components and flares}
   \end{figure}

   \placefigure{correlations between components and flares}

   To verify whether this anti-correlation is real or a consequence of variable boosting, it is necessary to make the 
   analysis in the comoving reference frame. As jet components are moving with relativistic velocities, the frequency of the emitted 
   radiation in the comoving frame $\nu^\prime$ and the frequency $\nu$ measured in the observer's reference frame are 
   related by $\nu=\delta \nu^\prime$. Besides, because time dilation, $\Delta T_{\mathrm{max}}(\nu)$ in the observer's 
   framework is related to $\Delta T_{\mathrm{max}}^\prime(\nu^\prime)$, defined in the comoving frame as:

   \begin{eqnarray}
      \Delta T_{\mathrm{max}}^\prime(\nu^\prime) = \Delta T_{\mathrm{max}}^\prime(\nu/\delta) = \delta \Delta T_{\mathrm{max}}(\nu)
   \end{eqnarray}

   But what we want is a relation between $\Delta T_{\mathrm{max}}^\prime$ and $\Delta T_{\mathrm{max}}$ always at 
   same comoving frequency, for example at frequency $\nu$. Assuming that each component propagates with constant velocity 
   in the comoving reference frame, \citet{abra01} showed that:

   \begin{eqnarray}
      \Delta T_{\mathrm{max}}^\prime(\nu/\delta) = \delta^{1/b} \Delta T_{\mathrm{max}}^\prime(\nu)
   \end{eqnarray}
   \\where the parameter $b$ depends on the dominating energy loss process \citep{mage85}. For instance, 
   for jet components in the adiabatic-loss stage, $b$ is given by:

   \begin{eqnarray}
       b = \frac{2(2s+1)+3a(s+2)}{3(s+4)}
   \end{eqnarray}

   \begin{eqnarray}
       s = 2\alpha_{\mathrm{r}}+1
   \end{eqnarray}
   \\where $\alpha_{\mathrm{r}}$ is the spectral index at radio wavelengths, and the parameter $a$ depends on 
   the variation of the magnetic field with the distance $y^\prime$ in the jet's reference frame 
   [$B(y^\prime) \propto y^{\prime-a}$]. For $a=1$, $B_\perp \approx B_\parallel$ ($B_\perp$ and $B_\parallel$ 
   are respectively the component of the magnetic field oriented perpendicularly and parallel to the jet axis), 
   while for $a=2$, $B_\parallel \gg B_\perp$.

   Substituting equation (15) in equation (14), we have that $\Delta T_{\mathrm{max}}$ measured in the observer's reference 
   frame and in the comoving frame at same frequency $\nu$ are related by:

   \begin{eqnarray}
      \Delta T_{\mathrm{max}}(\nu) = \delta^{1/b-1} \Delta T_{\mathrm{max}}^\prime(\nu)
   \end{eqnarray}

   In Cols. 4 and 5 of Table 3, we present respectively the Doppler factor $\delta_{\mathrm{0}}$ at the 
   formation epoch $t_{\mathrm{0}}$ and the measured spectral index $\alpha_{\mathrm{r}}$. Assuming that 
   components are in the adiabatic phase and $a=1$, we obtained $b$ through equation (16) (Col. 6 of Table 3). 
   With these parameters and equation (18), we determined $\Delta T_{\mathrm{max}}^\prime(\nu)$, displayed in 
   the Col. 7 of Table 3.

   The transformation of the optical flux density to its value in the the comoving reference frame depends on the 
   nature of the flare. If it is associated with some kind of instability in the accretion disk, such as perturbations 
   due to the passage of the secondary black hole through the disk \citep{rom00}, then the flux density 
   of the flares would not be boosted. As it can be seen in Fig. 4b, the anti-correlation between optical flux density 
   and $\Delta T_{\mathrm{max}}^\prime(\nu)$ is maintained, indicating that the stronger the flare is, the faster the 
   propagation of the perturbations in the disk is.

   Other possibility is that the optical flares are produced in the jet in the early stages of the shock evolution. 
   In this case, relativistic corrections are needed and: 

   \begin{eqnarray}
      S_{\mathrm{B}}=\delta^{3+\alpha_{\mathrm{op}}}S_{\mathrm{B}}^\prime
   \end{eqnarray}
   \\where $S_{\mathrm{B}}^\prime$ is the {\it B}-band flux density of the flare measured in the comoving frame. The 
   values of $S_{\mathrm{B}}^\prime$ are shown in the last column of Table 3.

   In Fig. 4c, we plotted $S_{\mathrm{B}}^\prime$ in terms of $\Delta T_{\mathrm{max}}^\prime(\nu)$. Again, we adopted 
   $\alpha_{\mathrm{opt}}=1.66$ \citep{hag96}. Clearly the remarkable anti-correlation seen 
   in Figs. 4a and b disappears completely, indicating that if the optical flares are short-lived phases of the 
   evolution of superluminal components, the anti-correlation is only a consequence of boosting. However, if the optical 
   flares originate from jet components, we would expect the anti-correlation between maximum flux density in the 
   observer's reference frame and $\Delta T_{\mathrm{max}}(\nu)$ to be observed also at radio frequencies. Since no 
   correlation was found between $\Delta T_{\mathrm{max}}(\nu)$ and maximum flux density at 10.7 or 22 GHz, the 
   scenario where the optical flares are produced in the accretion disk is favoured. It is important to 
   emphasize that the results obtained in this section are not altered if we had considered $a=2$ in our calculations.

\subsection{Polarimetric observations and precession model}

   Polarimetric interferometry of 3C\,345 \citep{bro94,lep95,tayl98,ros00} has shown that the electric 
   field orientation and the fractional polarization vary along its jet. \citet{bro94}, observing at 5 GHz, 
   found that the magnetic field is almost aligned with the local jet direction, except at the position of the 
   superluminal features. From observations at 22 GHz, \citet{loze96} and \citet{ros00} proposed the existence 
   of a transition region where the magnetic field changes its orientation 
   from transverse to longitudinal. In the inner parts, strong shocks would dominate, favouring magnetic field 
   perpendicular to the jet axis, while shocks in the outer region would be weaker, and they would not have too much 
   influence on the magnetic field orientation. The increase of the fractional polarization with distance along 
   the jet would be the result of the superposition, close to the core, of components with different electric field orientations 
   \citep{ros00}.

   If there were alignment between the magnetic field and the underlying jet, as in the case of 3C\,345, 
   jet inlet precession would introduce fast time variations in the polarization angle, specially when the 
   viewing angle is small. In fact, at the epochs for which the viewing angle is smallest, our model predicts 
   variations in position angle of 60 degrees during an interval as short as 2.6 yr (Fig. 2c). The vectorial 
   superposition of the electric fields associated with jet components formed during these epochs may lead 
   to the cancellation of the total polarized flux and consequent decrease in the fractional polarization 
   near the core. For larger distances, the differences in apparent velocities between the different components 
   would decrease the superposition and the fractional polarization would be higher.

\subsection{Physical parameters of a possible super-massive black hole binary system}

   The precession period of 10.1 yr assumed in our model is certainly inconsistent with the period of 
   1000 yr calculated by \citet{lu92} using the Lense-Thirring effect \citep{leth18}, for which jet 
   precession is due to the misalignment between the angular momenta of the accretion disk and 
   of a Kerr black hole. The precession of the accretion disk tidally induced by a secondary black hole in 
   a  black hole binary system seems to be a probable precession mechanism \citep{katz80,katz97,rom00}. 
   Assuming that last interpretation is correct for 3C\,345, we can estimate the physical parameters of the binary system.

   Let us consider that the primary and secondary black holes, with masses $M_{\mathrm{p}}$ and $M_{\mathrm{s}}$ 
   respectively, are separated by a distance $r_{\mathrm{ps}}$. From Kepler's third law, we can relate $r_{\mathrm{ps}}$ 
   to the orbital period of the secondary around the primary black hole $P_{\mathrm{ps}}$:

   \begin{eqnarray}
      r_{\mathrm{ps}}^3 = \frac{\mathrm{G}M_{\mathrm{tot}}}{4\pi^2}P_{\mathrm{ps}}^2
   \end{eqnarray}
   \\where G is the gravitational constant and $M_{\mathrm{tot}}$ is the sum of the masses of the two 
   black holes. 
   
   In the observer's reference frame, the orbital period $P_{\mathrm{ps}}^{\mathrm{obs}}$ will be:

   \begin{eqnarray}
      P_{\mathrm{ps}}^{\mathrm{obs}} = (1+z)P_{\mathrm{ps}}
   \end{eqnarray}

   If the orbit of the secondary is non-coplanar with the accretion disk, torques could be induced in the inner parts of 
   the accretion disk, producing its precession. Considering that the outer radius of the precessing part of the disk 
   is $r_{\mathrm{d}}$, \citet{pate95} and \citet{larw97} calculated its precession period $P_{\mathrm{d}}$ in 
   terms of the masses of the black holes:

   \begin{eqnarray}
     \frac{2\pi}{P_{\mathrm{d}}}(1+z) = -\frac{3}{4}\left(\frac{7-2n}{5-n}\right) \frac{GM_{\mathrm{s}}}{r_{\mathrm{ps}}^3}\frac{r_{\mathrm{d}}^2}{\sqrt{GM_{\mathrm{p}}r_{\mathrm{d}}}}\cos\theta
   \end{eqnarray}
   \\where $n$ is the politropic index of the gas (e.g., $n=3/2$ for a non-relativistic gas and $n=3$ for 
   the relativistic case) and $\theta$ is the inclination of the orbit of the secondary with respect to the plane of 
   the disk. 

   As it is believed that accretion disks are the main candidates for feeding jets,  they are expected to form 
   coupled systems (e.g., \citealt{dobi96,dobi02}). In this case, the jet would precess at the same 
   rate than the disk ($P=P_{\mathrm{d}})$, forming a precession cone with half-opening angle equal to the angle 
   of orbit inclination ($\Omega=\theta$). Thus, equation (22) becomes:

   \begin{eqnarray}
     \frac{2\pi}{P}(1+z) = -\frac{3}{4}\left(\frac{7-2n}{5-n}\right) \frac{G(M_{\mathrm{tot}}-M_{\mathrm{p}})}{\sqrt{GM_{\mathrm{p}}}}\left(\frac{\sqrt{r_{\mathrm{d}}}}{r_{\mathrm{ps}}}\right)^3\cos\Omega
   \end{eqnarray}
   \\where $M_{\mathrm{s}}$ was replaced by $M_{\mathrm{tot}}-M_{\mathrm{p}}$.

   From equation (23), $r_{\mathrm{d}}$ can be obtained in terms of $M_{\mathrm{p}}$ and $M_{\mathrm{tot}}$ by: 

   \begin{eqnarray}
      r_{\mathrm{d}} = \left[-\frac{8\pi}{3}\left(\frac{5-n}{7-2n}\right)\frac{(1+z)}{P\cos\Omega}\frac{r_{\mathrm{ps}}^3}{\sqrt{GM_{\mathrm{tot}}}}\right]^{2/3}\frac{x_{\mathrm{p}}^{1/3}}{\left(1-x_{\mathrm{p}}\right)^{2/3}}
   \end{eqnarray}
   \\where $x_{\mathrm{p}}=M_{\mathrm{p}}/M_{\mathrm{tot}}$. However, equation (24) is valid only if the disk precesses 
   like a rigid-body, implying that $r_{\mathrm{d}}$ must be appreciably smaller than $r_{\mathrm{ps}}$ \citep{pate95}. 
   Thus, if $M_{\mathrm{tot}}$ and $P_{\mathrm{ps}}$ are previously known, this condition can be used to put an upper 
   limit to the mass of the primary black hole, as well as a lower limit to the secondary mass.

   Using the velocity dispersion obtained from the H$\beta$ line-width and the optical continuum luminosity, 
   \citet{gu01} found a mass of $8\times 10^9$ M$_{\sun}$ for the central object in 3C\,345, which it is 
   assumed to be the value of $M_{\mathrm{tot}}$ in this work. In Table 4, we present the separation between the
   two components of the binary system and upper (lower) limits for the mass of the primary (secondary) black 
   hole for several orbital periods. We show in Fig. 5 the outer radius of the precession disk, as given by 
   equation (24), as a function of $x_{\mathrm{p}}$.

   We can rule out orbital periods $P_{\mathrm{ps}}^{\mathrm{obs}}<4$ yr because, in that case, the time-scale 
   for losses due to gravitational radiation is smaller than about 500 yr (e.g., \citealt{beg80,shte83}), which 
   would lead to significant changes in the orbit of the secondary and consequently to non-periodic disk precession. 
   In addition, from Table 4 we should not consider  periods $P_{\mathrm{ps}}^{\mathrm{obs}}>6.1$ yr since
   they correspond to a mass for the secondary 
   black hole  larger than that of the primary, which seems to be physically unlikely. 
   For these reasons, the possible ranges of  primary and secondary masses are 
   $4\times 10^9 \;{\mathrm M}_{\sun}\leq M_{\mathrm{p}} \leq 5\times 10^9$ M$_{\sun}$ and
   $3\times 10^9 \;{\mathrm M}_{\sun}\leq M_{\mathrm{s}} \leq 4\times 10^9$ M$_{\sun}$, respectively.
   
   According to \citet{rom00}, the passage of the secondary black hole through the primary accretion disk can produce 
   density waves and induce shocks in the jet, leading to the formation of the superluminal components. Since the 
   secondary black hole passes through the disk twice per orbit, the maximum efficiency is obtained when each passage 
   produces at least one jet component. Assuming that only one component per passage is formed, 
   the interval between successive ejections of superluminal components, measured 
   in the observer's reference frame $P_{\mathrm{ej}}^{\mathrm{obs}}$, corresponds to $P_{\mathrm{ps}}^{\mathrm{obs}}/2$. 
   If not all the interactions are able to generate a jet component, the observed periodicity would be a multiple of 
   half the orbital period ($P_{\mathrm{ej}}^{\mathrm{obs}}=\upsilon P_{\mathrm{ps}}^{\mathrm{obs}}/2$, with $\upsilon=1,2,3,...$). 
   From the emergence epochs of the jet components, we observe that the median ejection interval is 2.6 yr. 
   Assuming that this interval corresponds to the maximum ejection efficiency, $P_{\mathrm{ps}}^{\mathrm{obs}}=5.2$ yr, 
   similar to the period of 5.1 yr found in the 14.5-GHz and {\it B}-band light curves \citep{zha98,kel03}. Moreover, 
   if the 2.6-yr ejection period corresponds to twice the half period, we would obtain a full period of less than 4 yr, 
   incompatible with the condition found for stable precession mentioned above.

   On the other hand, if each passage of the secondary object through the 
   accretion disk generates multiple components $N_{c}$, the relation between $P_{\mathrm{ej}}^{\mathrm{obs}}$ 
   and $P_{\mathrm{ps}}^{\mathrm{obs}}$ is $P_{\mathrm{ej}}^{\mathrm{obs}}=\Upsilon P_{\mathrm{ps}}^{\mathrm{obs}}/2$, 
   where $\Upsilon=\upsilon/N_{c}$. 
   Even in the simplest case where $N_{c}=2$, and  assuming again $P_{\mathrm{ej}}^{\mathrm{obs}}=2.6$ yr and 
   $\upsilon=1$, we obtained 
   $P_{\mathrm{ps}}^{\mathrm{obs}}=10.4$ yr. However, this value for the orbital period corresponds again to a
    binary system where the secondary black hole 
   is more massive than the primary, so we can also rule out this possibility. Therefore, it is reasonable to conclude 
   that on average, each passage of the secondary through the accretion disk produces one superluminal component.

\begin{figure}
      {\plotone{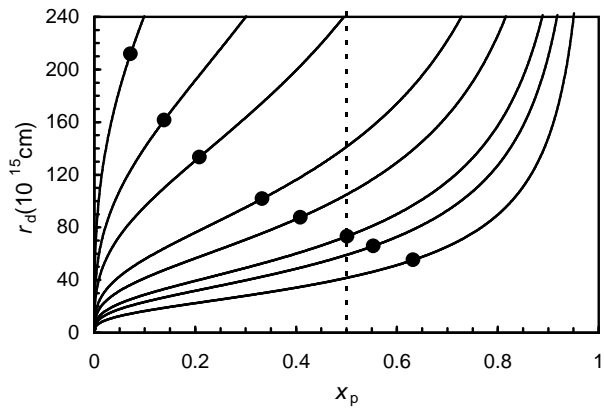}}
      \caption{Outer radius of the precessing disk as a function of $M_{\mathrm{p}}/M_{\mathrm{tot}}$. Each 
               full line corresponds to a different orbital period, increasing from right to left (see Table 4). 
			   The big 
               circles show the values at which the outer radius of the precessing disk equals
			   the separation between the two black holes. The dashed line marks the position 
               in which the masses of the primary and secondary black holes are equal.   
              }
         \label{Outer radius of the precessing disk}
   \end{figure}

   \placefigure{Outer radius of the precessing disk}
   \placetable{tbl-4}

\section{Conclusions}

   In this work, we used data available in the literature involving the parsec-scale radio structure, 
   which comprise almost twenty years of monitoring of 3C\,345. We assumed that all superluminal features 
   have quasi-ballistic motions in the inner regions ($r<1$ mas). 

   We showed that, close to the core, the trajectories of the superluminal components can be 
   explained quite well if the jet is precessing with a period of 10.1 yr, one of the {\it B}-band periodicities 
   reported by \citet{zha98}. As a consequence of this, we found a correlation between 
   time variation of the Doppler factor and the optical light curve, indicating that probably the boosted 
   optical emission from the underlying jet is responsible for the {\it B}-band long-term periodic variability. 
   An upper limit of 7 nJy for the intrinsic flux density of the underlying jet has been 
   estimated in this work. Lower limits for the Doppler factor calculated previously from X-ray observations 
   \citep{unw83,unw97} were used as constrains in our model.

   In this work, we also analyzed the influence of jet precession in the observed position of the optically 
   thick core at different frequencies \citep{blko79,loba98}. We found that precession introduces a time 
   modulation in the shift corrections, as well as in the direction in which the corrections should be applied. 
   However, core position shifts calculated in this work did not exceed 0.31 mas.

   Considering that jet precession in 3C\,345 is driven by a secondary super-massive black hole in a non-coplanar 
   orbit around the primary accretion disk and using the total mass of the two black holes derived from the H$\beta$ 
   line-width and optical continuum luminosity \citep{gu01}, we estimated the masses of the primary and
   secondary black holes as 
   $4\times 10^9 \;{\mathrm M}_{\sun}\leq M_{\mathrm{p}} \leq 5\times 10^9$ M$_{\sun}$ and
   $3\times 10^9 \;{\mathrm M}_{\sun}\leq M_{\mathrm{s}} \leq 4\times 10^9$ M$_{\sun}$, respectively.
   In addition, we found that the distance  $r_{\mathrm ps}$ between
   them should be $5.5\times 10^{16}\; {\mathrm {cm}} \leq r_{\mathrm {ps}}  \leq 7.3\times 10^{16}$ cm 
   and their orbital period $2.5\; {\mathrm {yr}} \leq P_{\mathrm {ps}} \leq 3.8$ yr.

   The observed dispersion in the apparent sizes of jet components could be attributed to jet precession and 
   interactions between jet and environment matter. It is corroborated by sudden variations in the radio 
   flux density of the knots when there are changes in their propagation direction. Furthermore, those interactions 
   seem to be responsible for gentle bending seen at distances larger than 1 mas, which could explain also 
   the difference between jet orientation at parsec- and kiloparsec-scales.

   We were able to associate the emergence of jet components to the occurrence of flares in the optical band 
   stronger than about 1.5 mJy. We also found an unexpected relation between the flare flux density at optical 
   wavelengths and the elapsed time between the epochs of superluminal component ejection and occurrence of 
   maximum flux density at radio frequencies, which led us to the conclusion that these optical flares are 
   originated in the accretion disk and not in the shock waves seen as superluminal features.
      
   Finally, the complex polarimetric structure of 3C\,345 in parsec-scales \citep{bro94,lep95,tayl98,ros00} may 
   be explained by our precession model, when the superposition of jet components with different position angles 
   is considered.  

\acknowledgments

   This work was supported by the Brazilian Agencies FAPESP (Proc. 99/10343-3), CNPq and FINEP. We would like 
   to thank the anonymous referee for careful reading of the manuscript and for useful comments and suggestions.

\appendix

\section{Opacity effects on core-component distance in a precessing jet}

   The kinematic properties of the superluminal components in the jet of 3C\,345 depend on the accurate determination 
   of the core-component separation. As pointed out previously (e.g., \citealt{blko79,loba96,loba98}), the absolute core 
   position $r_{\mathrm{core}}$ depends inversely on the frequency when the core is optically thick, which introduces a 
   shift in the core-component separation. Following \citet{blko79}, we can write the absolute core position as:  

   \begin{eqnarray}
      r_{\mathrm{core}}(\nu) = \frac{4.56\times 10^{-12}(1+z)}{D_{\mathrm{L}}\gamma^2 k_{\mathrm{e}}^{1/3}\psi \sin\phi} \left[\frac{L_{\mathrm{syn}}\sin\phi}{\beta(1-\beta\cos\phi)\ln(\gamma_{\mathrm{max}}/\gamma_{\mathrm{min}})}\right]^{2/3}\nu^{-1} (\mathrm{mas})
   \end{eqnarray}
   \\where $z$ is the redshift, $D_{\mathrm{L}}$ is the luminosity distance (in units of parsec), 
   $k_{\mathrm{e}}$ is a constant ($k_{\mathrm{e}}\leq 1$; \citealt{blko79}), $L_{\mathrm{syn}}$ 
   is the integrated synchrotron luminosity (in units of erg s$^{-1}$), while $\gamma_{\mathrm{max}}$ 
   and $\gamma_{\mathrm{min}}$ are related respectively to the upper and lower limits of the energy 
   distribution of the relativistic jet particles. The quantities $\psi$ and $\nu$ are respectively 
   the observed aperture angle of the jet (in radians) and the frequency (in Hz); the former is 
   related to the intrinsic jet aperture angle $\psi^\prime$ through (e.g., \citealt{mut90}):

   \begin{eqnarray}
      \tan(\psi/2) = \tan(\psi^\prime/2)\cot\phi 
   \end{eqnarray}  

   The core position shift $\Delta r_{\mathrm{core}}$ between frequencies $\nu_1$ and $\nu_2$ 
   ($\nu_2\geq\nu_1$) is given by:

   \begin{eqnarray}
      \Delta r_{\mathrm{core}}(\nu_1,\nu_2) = \frac{4.56\times 10^{-12}(1+z)}{D_{\mathrm{L}}\gamma^2 k_{\mathrm{e}}^{1/3}\psi \sin\phi} \left[\frac{L_{\mathrm{syn}}\sin\phi}{\beta(1-\beta\cos\phi)\ln(\gamma_{\mathrm{max}}/\gamma_{\mathrm{min}})}\right]^{2/3}\frac{(\nu_2-\nu_1)}{\nu_1\nu_2} (\mathrm{mas})
   \end{eqnarray}

   Note that if we substitute in equation (A3) a simpler version of equation (A2), $\psi\approx \psi^\prime\csc\theta$, 
   we obtain equation (11) given in \citet{loba98}.

   We can see that equation (A3) depends on the angle between the jet and line of sight; in the case 
   of a jet which is precessing, this angle is a function of time, what obviously introduces a 
   time dependency in $\Delta r_{\mathrm{core}}$. On the other hand, the shifts 
   in the core-component separations do not occur in a fixed direction, but they are oriented 
   according to the direction at which the jet inlet is pointed. As the jet inlet is not resolved by 
   observations, changes in its direction will reflect on changes in the position angle of the 
   core region. Thus, a jet component, located at a distance $r$ from the core and with a position 
   angle $\eta$, will have right ascension and declination offsets ($\Delta\alpha$ and $\Delta\delta$ 
   respectively) given by:

   \begin{mathletters}
    \begin{eqnarray}
      \Delta\alpha(\nu_1,\nu_2) = r(\nu_1)\sin[\eta(t_{\mathrm{obs}})]-\Delta r_{\mathrm{core}}(\nu_1,\nu_2)\sin[\eta_{\mathrm{c}}(t_{\mathrm{obs}})]\\ 
      \Delta\delta(\nu_1,\nu_2) = r(\nu_1)\cos[\eta(t_{\mathrm{obs}})]-\Delta r_{\mathrm{core}}(\nu_1,\nu_2)\cos[\eta_{\mathrm{c}}(t_{\mathrm{obs}})]
    \end{eqnarray}
   \end{mathletters}
   \\where $\eta_{\mathrm{c}}$ is the position angle of the core at the epoch $t_{\mathrm{obs}}$ in which 
   observation is acquired. Once the precession model parameters are known, we are able to determine the 
   second term of the equations (A4a) and (A4b) and correct the component position by core opacity effects; 
   using them, we can determine the corrected core-component distance $r_{\mathrm{corr}}$ 
   through:

   \begin{eqnarray}
      r_{\mathrm{corr}}(\nu_1) = \sqrt{r(\nu_1)^2+\Delta r_{\mathrm{core}}(\nu_1,\nu_2)^2-2r(\nu_1)\Delta r_{\mathrm{core}}(\nu_1,\nu_2)\cos(\eta-\eta_{\mathrm{c}})}
   \end{eqnarray}

   Note that if $\Delta r_{\mathrm{core}}=0$, $r_{\mathrm{corr}}=r$. Other particular case 
   is found when there is alignment between the position angles of the core and of the jet 
   component ($\eta=\eta_{\mathrm{c}}$), such that 
   $r_{\mathrm{corr}}=r-\Delta r_{\mathrm{core}}$.

\clearpage


\twocolumn

\onecolumn

\twocolumn

\clearpage
\onecolumn

\twocolumn

\clearpage


\begin{deluxetable}{cccccc}
\tabletypesize{\scriptsize}
\tablecaption{Parameters of the superluminal components of 3C\,345. \label{tbl-1}}
\tablewidth{0pt}
\tablehead{
\colhead{Component} & \colhead{$t_{\mathrm{0}}$ (yr)} & \colhead{$\mu$ (mas yr$^{-1}$)} &
\colhead{$h\beta_{\mathrm{obs}}$} & \colhead{$\eta$ ($\degr$)} & \colhead{OF}
}
\startdata
C3a               & 1977.5~$\pm$~1.5 & 0.355~$\pm$~0.025 & 8.7~$\pm$~0.4 &  -98~$\pm$~6  & yes\\
C4                & 1980.1~$\pm$~1.3 & 0.225~$\pm$~0.015 & 5.5~$\pm$~0.3 & -120~$\pm$~8  & yes\tablenotemark{a}\\
C5                & 1982.3~$\pm$~1.0 & 0.200~$\pm$~0.030 & 4.9~$\pm$~0.6 & -107~$\pm$~8  & yes\\
C6                & 1987.5~$\pm$~1.0 & 0.347~$\pm$~0.030 & 8.5~$\pm$~0.6 &  -98~$\pm$~8  & no\tablenotemark{b}\\
C7                & 1991.0~$\pm$~1.0 & 0.208~$\pm$~0.025 & 5.1~$\pm$~0.5 &  -97~$\pm$~6  & yes\\
C8                & 1993.0~$\pm$~1.0 & 0.196~$\pm$~0.030 & 4.8~$\pm$~0.6 & -110~$\pm$~6  & yes\tablenotemark{a}\\
 \enddata


\tablecomments{$t_{\mathrm{0}}$ is the formation epoch, $\mu$ and $h\beta_{\mathrm{obs}}$ are, 
respectively, the proper motion and the apparent velocity in units of light speed $c$, 
and $\eta$ is the ejection angle. Col.OF indicates whether or not there is some optical 
flare observed in the {\it B}-band associated to their ejections.}


\tablenotetext{a}{Considering the uncertainty of $\mathrm{t_{0}}$.}
\tablenotetext{b}{There is not any optical observation in the {\it B}-band available at this epoch.}

\end{deluxetable}
%
%


\begin{deluxetable}{ccccc}
\tabletypesize{\scriptsize}
\tablecaption{Parameters of the precession model for the parsec-jet of 3C\,120. \label{tbl-2}}
\tablewidth{0pt}
\tablehead{
\colhead{$P$ (yr)}\tablenotemark{a} & \colhead{$\gamma$} & \colhead{$\Omega$ ($\degr$)} & 
\colhead{$\phi_{\mathrm{0}}$ ($\degr$)} & \colhead{$\eta_{\mathrm{0}}$ ($\degr$)}
}
\startdata
10.1~$\pm$~0.8\tablenotemark{b} & 12.5~$\pm$~0.5 & 1.3~$\pm$~0.5 & 2.6~$\pm$~0.5 & -96~$\pm$~5\\
 \enddata
\tablenotetext{a}{Measured in the framework fixed at the observer.}
\tablenotetext{b}{The quoted error refers to that given by \citet{zha98}.}
\end{deluxetable}
%
%

\clearpage


\begin{deluxetable}{ccccccccc}
\tabletypesize{\scriptsize}
\tablecaption{Optical flares and radio evolution of the superluminal components of 3C\,345. \label{tbl-3}}
\tablewidth{0pt}
\tablehead{
\colhead{Component} & \colhead{$S_{\mathrm{op}}$ (mJy)} & \colhead{$\Delta T_{\mathrm{max}}$ (yr)} & 
\colhead{$\delta_{\mathrm{0}}$} & \colhead{$\alpha_{\mathrm{r}}$} & \colhead{$b$\tablenotemark{a}} & 
\colhead{$\Delta T_{\mathrm{max}}^\prime$ (yr)\tablenotemark{a}} & \colhead{$S_{\mathrm{op}}^\prime$ ($\mu$Jy)}
}
\startdata
C3                & 1.69~$\pm$~0.12    & 3.44~$\pm$~0.27\tablenotemark{b} & 14.79  & 0.50~$\pm$~0.35 & 1.22~$\pm$~0.10 & 5.6~$\pm$~1.1 & 6.0~$\pm$~1.3\\
C3a               & 2.01~$\pm$~0.32    & 1.94~$\pm$~0.37\tablenotemark{b} & 14.65  & 0.76~$\pm$~0.10 & 1.31~$\pm$~0.03 & 3.7~$\pm$~0.7 & 7.4~$\pm$~1.2\\
C4                & 1.76~$\pm$~0.60    & 2.32~$\pm$~0.18\tablenotemark{c} & 18.72  & 0.55~$\pm$~0.35 & 1.24~$\pm$~0.15 & 4.1~$\pm$~1.2 & 2.1~$\pm$~0.7\\
C5                & 2.38~$\pm$~0.22    & 1.46~$\pm$~0.16\tablenotemark{c} & 22.86  & 0.80~$\pm$~0.15 & 1.32~$\pm$~0.09 & 3.1~$\pm$~0.6 & 1.1~$\pm$~0.1\\
C7                & 2.44~$\pm$~0.38    & 1.44~$\pm$~0.19\tablenotemark{c} & 21.35  & 0.61~$\pm$~0.16 & 1.26~$\pm$~0.06 & 2.7~$\pm$~0.5 & 1.6~$\pm$~0.2\\
 \enddata
\tablecomments{$S_{\mathrm{opt}}$ is the flux density of the {\it B}-band optical flares. $\Delta T_{\mathrm{max}}$ 
                 is the interval between formation of the jet component and its maximum emission at a given radio 
                 frequency (see text). $\delta_{\mathrm{0}}$ is the Doppler factor at the emergence epoch, 
                 $\alpha_{\mathrm{r}}$ is the spectral index at radio wavelengths and parameter $b$, defined by equation (16). 
                 Finally, $\Delta T_{\mathrm{max}}^\prime$ and $S_{\mathrm{B}}^\prime$ are respectively the interval 
                 between formation of the jet component and its maximum emission and the {\it B}-band flux density of the flares 
                 in the comoving frame.}
\tablenotetext{a}{Considering $a=1$;}
\tablenotetext{b}{At 10.7 GHz;}
\tablenotetext{c}{At 22 GHz.}

\end{deluxetable}
%
%


\begin{deluxetable}{ccccc}
\tabletypesize{\scriptsize}
\tablecaption{Parameters of a possible black hole binary system in the inner parts of 3C\,345. \label{tbl-4}}
\tablewidth{0pt}
\tablehead{
\colhead{$P_{\mathrm{ps}}^{\mathrm{obs}}$ (yr)} &
\colhead{$P_{\mathrm{ps}}$ (yr)} & \colhead{$r_{\mathrm{ps}}$ (cm)} & 
\colhead{$M_{\mathrm{p}}$ (M$_{\sun}$)\tablenotemark{a}} & 
\colhead{$M_{\mathrm{s}}$ (M$_{\sun}$)
\tablenotemark{b}}
}
\startdata
 4.0 &  2.5 & $5.5\times10^{16}$ & $5.0\times10^9$ & $3.0\times10^9$\\
 5.2 &  3.3 & $6.6\times10^{16}$ & $4.4\times10^9$ & $3.6\times10^9$\\
 6.1 &  3.8 & $7.3\times10^{16}$ & $4.0\times10^9$ & $4.0\times10^9$\\
 8.0 &  5.0 & $8.8\times10^{16}$ & $3.2\times10^9$ & $4.8\times10^9$\\
10.0 &  6.3 & $1.0\times10^{17}$ & $2.6\times10^9$ & $5.4\times10^9$\\
15.0 &  9.4 & $1.3\times10^{17}$ & $1.6\times10^9$ & $6.4\times10^9$\\
20.0 & 12.6 & $1.6\times10^{17}$ & $1.0\times10^9$ & $7.0\times10^9$\\
30.0 & 18.8 & $2.1\times10^{17}$ & $5.6\times10^8$ & $7.4\times10^9$\\
 \enddata
\tablenotetext{a}{Upper limit;}
\tablenotetext{b}{Lower limit.}
\end{deluxetable}

\end{document}